\documentclass[showpacs,prb,twocolumn]{revtex4}
\usepackage{amsfonts}
\usepackage{amsmath}
\usepackage{amssymb}
\usepackage{graphicx}
\begin{document}

\title{The SU(3) bosons and the spin nematic state on the spin-1 bilinear-biquadratic
triangular lattice}
\author{Peng Li$^{1}$, Guang-Ming Zhang$^{2}$, and Shun-Qing Shen$^{1}$, }
\affiliation{$^{1}$Department of Physics, and Center for Theoretical and Computational
Physics, The University of Hong Kong, Pokfulam Road, Hong Kong, China}
\affiliation{$^{2}$Department of Physics and Center for Advanced Study, Tsinghua
University, Beijing 100084, China.}

\begin{abstract}
A bond-operator mean-field theory in the SU(3) bosons representation is
developed to describe the antiferro-nematic phase of the spin-1
bilinear-biquadratic model. The calculated static structure factors reveal
delicately that the antiferro-nematic state may exhibit both the ferro- and
antiferro-quadruple long-range orders, which is reminiscent of the
ferrimagnets or the canted antiferromagnets. This result may influence the
spin wave theory concerned with this phase. Possible relevance of this
unconventional state to the quasi-two-dimensional triangular material
NiGa$_{2}$S$_{4}$ is addressed.

\end{abstract}
\date{\today}

\maketitle

\section{Introduction}

The spin-$1$ bilinear-biquadratic model (SBBM)%
\begin{equation}
H=J_{\varphi}\sum_{\left\langle ij\right\rangle }\left[  \cos\varphi
\ \mathbf{S}_{i}\cdot\mathbf{S}_{j}+\sin\varphi\ (\mathbf{S}_{i}%
\cdot\mathbf{S}_{j})^{2}\right] \label{H0}%
\end{equation}
was put forward long time ago \cite{Chen1971,Chen1973,Papanicolaou1988}, where
$\mathbf{S}_{i}$ is the spin-$1$ operator. In one dimension, the phase diagram
was established \cite{Chubukov,Schollwok1996,Fath,Lauchli-cond}, but there are
still some controversies \cite{Lauchli-cond,ZhangGM}. The phase diagram in two
dimensions and above may be simpler because of suppression of quantum
fluctuations. Generally speaking, there are two regimes exhibiting different
types of spin nematic orders: (1) the ferro-nematic phase for $-3\pi
/4<\varphi<-\pi/2$ (2) the antiferro-nematic phase for $\pi/4<\varphi<\pi/2$.
Recently the first regime with ferro-quadruple long-range order (LRO) attracts
much attention due to the fact that the Mott insulating state was realized in
a system of bosonic atoms in an optical lattice
\cite{Greiner2002,Yip2003,Imambekov2003}. Here we shall study the second
regime by a bond-operator mean-field theory in SU($3$) bosons representation.
The unconventional properties of this nematic state, such as the absence of
magnetic LRO and the gapless excitation, are quite instructive for explaining
recent experimental observations in NiGa$_{2}$S$_{4}$
\cite{Nakatsuji2005,Tsunetsugu,Lauchli2}.

In a framework of frustrated SU($N$) model, we expressed the SBBM in terms of
SU($3$) generators and proposed an associated bond-operator mean-field theory
in both bosonic and fermionic representations \cite{Zhang01,Li2004}. The
theory is a generalization of the widely used Schwinger-boson mean-field
theory (SBMFT) \cite{Auerbach1988}. The advantage of the theory is that we can
use it to study either the ordered or disordered phases. In this paper, we
shall use the bosonic theory to study the unconventional orders of the
antiferro-nematic states on the triangular lattice. It will be shown that the
ferro- and antiferro-quadruple LRO's may coexist at low temperature for the
quadruple operators, which is reminiscent of the ferrimagnets or the canted
antiferromagnets. And the uniform quadruple moments may keep nonzero at finite
temperatures. These two new features enrich our knowledge of the
antiferro-nematic state of this model. To show the relevance of this state to
the observations in NiGa$_{2}$S$_{4}$, we also calculate the physical
quantities, e.g. the ground energy, the specific heat, and the uniform
magnetic susceptibility. A similar theory with a different scheme had been
applied to the ferro-nematic phase by one of the authors in a previous work
\cite{ZhangGM}.

The paper is organized as follows. In Sec. II we introduce the SU(3) boson
representation for spin 1 system, and express the Hamiltonian of Eq.
(\ref{H0}) in terms of SU(3) generators. In Sec. III, we present the formalism
of the bond-operator mean-field theory in bosonic language. Then in Sec. IV,
we work out the mean-field equations and uncover some properties of the
antiferro-nematic phase. In Sec. V, we present discussions of our results.

\section{SU($3$) Bosons Representation}

In SBBM, each site has three states, $\left\vert m_{\alpha}\right\rangle $
with $m_{1}=-1,m_{2}=0$,$\ $and $m_{3}=+1$, according to the eigenvalues of
the $z$-component of spin, $S^{z}$. We reorganize the three states and
introduce three bosonic creation operators,
\begin{subequations}
\label{Qeigenstates}%
\begin{align}
b_{1}^{\dag}\left\vert 0\right\rangle  &  =\frac{1}{\sqrt{2}}\left(
\left\vert m_{1}\right\rangle -\left\vert m_{3}\right\rangle \right)  ,\\
b_{2}^{\dag}\left\vert 0\right\rangle  &  =\frac{i}{\sqrt{2}}\left(
\left\vert m_{1}\right\rangle +\left\vert m_{3}\right\rangle \right)  ,\\
b_{3}^{\dag}\left\vert 0\right\rangle  &  =\left\vert m_{2}\right\rangle .
\end{align}
In terms of $b$ operators, the eight generators of SU($3$) group can be
expressed by three boson operators. They can be divided into two categories,
the spin operators
\end{subequations}
\begin{subequations}
\begin{align}
S_{i}^{x} &  =-i(b_{i2}^{\dag}b_{i3}-b_{i3}^{\dag}b_{i2}),\\
S_{i}^{y} &  =-i(b_{i3}^{\dag}b_{i1}-b_{i1}^{\dag}b_{i3}),\\
S_{i}^{z} &  =-i(b_{i1}^{\dag}b_{i2}-b_{i2}^{\dag}b_{i1}),
\end{align}
and the quadrupole operators (second-order spin moments)
\end{subequations}
\begin{subequations}
\label{quadrupole operators}%
\begin{align}
Q_{i}^{\left(  0\right)  } &  =\left(  S_{i}^{z}\right)  ^{2}-\frac{2}%
{3}=\frac{1}{3}\left(  b_{i1}^{\dag}b_{i1}+b_{i2}^{\dag}b_{i2}-2b_{i3}^{\dag
}b_{i3}\right)  ,\\
Q_{i}^{\left(  2\right)  } &  =\left(  S_{i}^{x}\right)  ^{2}-\left(
S_{i}^{y}\right)  ^{2}=-\left(  b_{i1}^{\dag}b_{i1}-b_{i2}^{\dag}%
b_{i2}\right)  ,\\
Q_{i}^{xy} &  =S_{i}^{x}S_{i}^{y}+S_{i}^{y}S_{i}^{x}=-\left(  b_{i1}^{\dag
}b_{i2}+b_{i2}^{\dag}b_{i1}\right)  ,\\
Q_{i}^{yz} &  =S_{i}^{y}S_{i}^{z}+S_{i}^{z}S_{i}^{y}=-\left(  b_{i2}^{\dag
}b_{i3}+b_{i3}^{\dag}b_{i2}\right)  ,\\
Q_{i}^{zx} &  =S_{i}^{z}S_{i}^{x}+S_{i}^{x}S_{i}^{z}=-\left(  b_{i3}^{\dag
}b_{i1}+b_{i1}^{\dag}b_{i3}\right)  .
\end{align}
In this case the Hamiltonian, Eq. (\ref{H0}), can be expressed in terms of
these generators and has a form of the generalized frustrated SU($3$) model
\cite{Li2004},
\end{subequations}
\begin{equation}
H=\sum_{\left\langle ij\right\rangle }\Upsilon_{1}\left(  i,j\right)
+\sum_{\left\langle ij\right\rangle }\Upsilon_{2}\left(  i,j\right)  +\sum
_{i}\lambda_{i}\left(  \sum\limits_{\mu}b_{i\mu}^{\dag}b_{i\mu}-1\right)
,\label{H}%
\end{equation}
where%
\begin{equation}
\Upsilon_{1}\left(  i,j\right)  =J_{1}\sum_{\mu\nu}\mathcal{J}_{\nu}^{\mu
}(r_{i})\mathcal{J}_{\mu}^{\nu}(r_{j}),
\end{equation}%
\begin{equation}
\Upsilon_{2}\left(  i,j\right)  =-J_{2}\sum_{\mu\nu}\mathcal{J}_{\nu}^{\mu
}(r_{i})\mathcal{J}_{\nu}^{\mu}(r_{j}),
\end{equation}
with $J_{1}=J_{\varphi}\cos\varphi$ and $J_{2}=J_{\varphi}\left(  \cos
\varphi-\sin\varphi\right)  $. $\mathcal{J}_{\nu}^{\mu}(r_{i})=b_{i\mu}^{\dag
}b_{i\nu}$ are the generators of the SU($3$) group, and the Lagrangian
multipliers $\lambda_{i}$ are introduced to realize the single occupancy of
the bosons at each lattice site. The first term in Eq.(\ref{H}) possesses the
SU($3$) symmetry because the operator,
\begin{equation}
\sum_{\mu\nu}\mathcal{J}_{\nu}^{\mu}(r_{i})\mathcal{J}_{\mu}^{\nu}%
(r_{j})\equiv P_{ij},
\end{equation}
serves as the permutation operator, which swaps two quantum states at sites
$i$ and $j$,%
\begin{equation}
P_{ij}\left\vert i,\mu;j,\nu\right\rangle =\left\vert i,\nu;j,\mu\right\rangle
.
\end{equation}
The second term in Eq.(\ref{H}) breaks the SU($N$) symmetry on the triangular lattice.

\section{Mean-Field Theory}

\subsection{Decomposition Scheme}

Now we concentrate on the regime with $J_{1}>0$ and $J_{2}<0$. In the boson
representation, we introduce two types of bond operators
\begin{subequations}
\begin{align}
\Delta_{ij,\mu\nu} &  =b_{j\mu}b_{i\nu}-b_{j\nu}b_{i\mu},\ \left(  \mu
<\nu\right)  ,\\
\Xi_{ij,\mu\nu} &  =b_{j\mu}^{\dagger}b_{i\nu}-b_{j\nu}^{\dagger}b_{i\mu
},\ \left(  \mu<\nu\right)  ,
\end{align}
and the four-operator terms in the Hamiltonian Eq. (\ref{H}) can be written
as
\end{subequations}
\begin{subequations}
\begin{align}
\Upsilon_{1}\left(  i,j\right)   &  =-J_{1}\sum_{\mu<\nu}\Delta_{ij,\mu\nu
}^{\dagger}\Delta_{ij,\mu\nu}+J_{1},\\
\Upsilon_{2}\left(  i,j\right)   &  =-\left\vert J_{2}\right\vert \sum
_{\mu<\nu}\text{:}\Xi_{ij,\mu\nu}^{\dagger}\Xi_{ij,\mu\nu}\text{:}+\left\vert
J_{2}\right\vert .
\end{align}
Notice the single occupancy constraint, $\sum\limits_{\mu}b_{i\mu}^{\dag
}b_{i\mu}=1$, is used when the expressions are deduced. Consider that the
model is isotropic, one can introduce two real mean-field parameters
\end{subequations}
\begin{subequations}
\label{mean fields for BBM}%
\begin{align}
\Delta &  =\left\langle \Delta_{ij,\mu\nu}\right\rangle =\left\langle
\Delta_{ij,\mu\nu}^{\dag}\right\rangle ,\\
\Xi &  =\left\langle \Xi_{ij,\mu\nu}\right\rangle =\left\langle \Xi_{ij,\mu
\nu}^{\dag}\right\rangle .
\end{align}
The prescribed mean fields represent the ultrashort-range correlations.
However, when the mean field equations are solved, some physical restrictions
should be fulfilled. For instance, the biquadratic term in the SBBM ($N=3$)
can be written as
\end{subequations}
\begin{equation}
\left(  \mathbf{S}_{i}\cdot\mathbf{S}_{j}\right)  ^{2}=-\sum_{\mu<\nu}%
\text{:}\Xi_{ij,\mu\nu}^{\dagger}\Xi_{ij,\mu\nu}\text{:}+1,\ \left(  \mu
,\nu=1,2,3\right)  .
\end{equation}
Since $\left(  \mathbf{S}_{i}\cdot\mathbf{S}_{j}\right)  ^{2}\geqslant0$, one
would obtain the restriction%
\begin{equation}
\Xi=\left\langle \Xi_{ij,\mu\nu}\right\rangle \leqslant\frac{1}{\sqrt{3}}.
\end{equation}
Our numerical result shows that this restriction is well satisfied.

\subsection{Mean-Field Equations}

\begin{figure}
[ptb]
\begin{center}
\includegraphics[
height=1.303in,
width=2.8889in
]%
{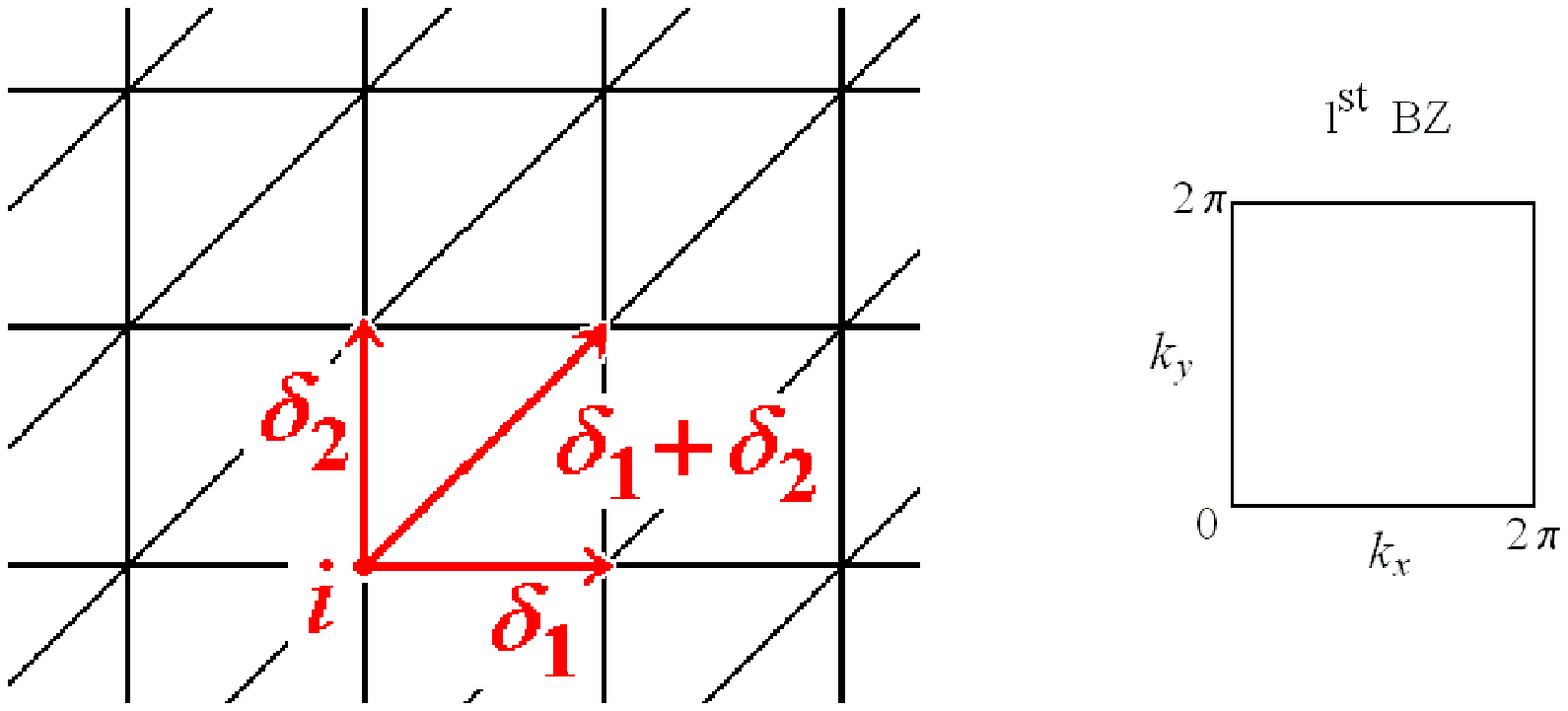}%
\caption{(Color online) The lattice we used in the calculation, which is
topologically equivalent to the triangular lattice given that the interactions
along directions, $\delta_{1},\delta_{2},$ and $\delta_{1}+\delta_{2}$, are
equal. On this lattice, the first Brillouin zone is a square with volume
$\left(  2\pi\right)  ^{2}$.}%
\label{triangular lattice}%
\end{center}
\end{figure}

We limit our calculation on the triangular lattice (see Fig.
\ref{triangular lattice}).

The Hubbard-Stratonovich transformation is performed to decouple the
Hamiltonian Eq. (\ref{H}) into a bilinear form%
\begin{align}
H &  =-J_{1}\Delta\sum_{\mu<\nu}\sum_{i,\delta>0}\left(  \Delta_{i,i+\delta
;\mu\nu}+\Delta_{i,i+\delta;\mu\nu}^{\dag}\right) \nonumber\\
&  -\left\vert J_{2}\right\vert \Xi\sum_{\mu<\nu}\sum_{i,\delta}\left(
\Xi_{i,i+\delta;\mu\nu}+\Xi_{i,i+\delta;\mu\nu}^{\dag}\right) \nonumber\\
&  +\frac{z}{2}N_{\Lambda}N\left(  J_{1}\Delta^{2}+\left\vert J_{2}\right\vert
\Xi^{2}\right)  +\lambda\sum_{i}\sum\limits_{\mu}b_{i\mu}^{\dag}b_{i\mu
}-\lambda N_{\Lambda},
\end{align}
where $\sum_{\delta>0}$ means summation over the nearest neighbours in the
positive directions of a given site, $N_{\Lambda}$ is the total number of
lattice sites, $z$ is the coordinate number of the lattice, e.g. $z=6$ for the
triangular lattice.

After performing the Fourier transform and introducing the Nambu spinor in the
momentum space,
\begin{equation}
\Phi_{k}^{\dagger}=\left(  b_{k,1}^{\dagger},b_{k,2}^{\dag},b_{k,3}^{\dagger
},b_{-k,1},b_{-k,2},b_{-k,3}\right)  .
\end{equation}
one can arrive at the mean-field Hamiltonian for the spin $S$ chain in\ a
compact form,%
\begin{equation}
H=\frac{1}{2}\sum_{k}\Phi_{k}^{\dagger}M_{k}\Phi_{k}+\varepsilon_{0},
\end{equation}
where
\begin{subequations}
\begin{align}
M_{k} &  =\lambda\sigma^{0}\otimes A_{0}+i\Delta_{k}\sigma^{x}\otimes
A_{1}+i\Xi_{k}\sigma^{0}\otimes A_{1},\\
A_{0} &  =\left(
\begin{array}
[c]{ccc}%
1 & 0 & 0\\
0 & 1 & 0\\
0 & 0 & 1
\end{array}
\right)  ,\ A_{1}=\left(
\begin{array}
[c]{ccc}%
0 & -1 & -1\\
1 & 0 & -1\\
1 & 1 & 0
\end{array}
\right)  ,\\
\Delta_{k} &  =2J_{1}\Delta\eta_{k},\Xi_{k}=2\left\vert J_{2}\right\vert
\Xi\eta_{k},\eta_{k}=\sum_{\delta>0}\sin k_{\delta}\label{DeltaCheta}\\
\varepsilon_{0} &  =\frac{3}{2}zN_{\Lambda}\left(  J_{1}\Delta^{2}+\left\vert
J_{2}\right\vert \Xi^{2}\right)  -\frac{5}{2}\lambda N_{\Lambda}.
\end{align}
By diagonalizing the Hamiltonian, we get three spectra,
\end{subequations}
\begin{subequations}
\label{spectra BBM boson}%
\begin{align}
\omega_{1}\left(  k\right)   &  =\lambda,\\
\omega_{2}\left(  k\right)   &  =\sqrt{\left(  \lambda-\sqrt{3}\Xi_{k}\right)
^{2}-\left(  \sqrt{3}\Delta_{k}\right)  ^{2}},\\
\omega_{3}\left(  k\right)   &  =\sqrt{\left(  \lambda+\sqrt{3}\Xi_{k}\right)
^{2}-\left(  \sqrt{3}\Delta_{k}\right)  ^{2}}.
\end{align}
Notice that the two spectra $\omega_{3}\left(  k\right)  $ and $\omega
_{2}\left(  k\right)  $ has a relation of $\omega_{3}\left(  -k\right)
=\omega_{2}\left(  k\right)  $. By optimization of the total free energy
\end{subequations}
\begin{equation}
F=\varepsilon_{0}-\frac{1}{\beta}\sum\limits_{k,\mu}\ln\left[  n_{B}\left(
\omega_{\mu}\right)  \left(  n_{B}\left(  \omega_{\mu}\right)  +1\right)
\right]  ,
\end{equation}
where $n_{B}\left(  \omega_{\mu}\right)  $ is the Boltzmann distribution
function, three mean-field equations are established
\begin{subequations}
\begin{gather}
2-n_{B}\left(  \lambda\right)  =\int\frac{d^{2}k}{\left(  2\pi\right)  ^{2}%
}\frac{1-\widetilde{\Xi}\ \eta_{k}}{\widetilde{\omega}_{2}\left(  k\right)
}\coth\frac{\beta\widetilde{\omega}_{2}\left(  k\right)  }{2},\\
\Delta=\frac{1}{3\sqrt{3}}\int\frac{d^{2}k}{\left(  2\pi\right)  ^{2}}%
\frac{\widetilde{\Delta}\ \eta_{k}^{2}}{\widetilde{\omega}_{2}\left(
k\right)  }\coth\frac{\beta\widetilde{\omega}_{2}\left(  k\right)  }{2},\\
\Xi=\frac{1}{3\sqrt{3}}\int\frac{d^{2}k}{\left(  2\pi\right)  ^{2}}%
\frac{\left(  1-\widetilde{\Xi}\ \eta_{k}\right)  \ \eta_{k}}{\widetilde
{\omega}_{2}\left(  k\right)  }\coth\frac{\beta\widetilde{\omega}_{2}\left(
k\right)  }{2},
\end{gather}
in which we have introduced dimensionless quantities for convenience of
calculation
\end{subequations}
\begin{subequations}
\begin{align}
\widetilde{\omega}_{\mu}\left(  k\right)   &  =\frac{\omega_{\mu}\left(
k\right)  }{\lambda},\\
\widetilde{\Delta} &  =\frac{2\sqrt{3}J_{1}\Delta}{\lambda},\\
\widetilde{\Xi} &  =\frac{2\sqrt{3}\left\vert J_{2}\right\vert \Xi}{\lambda}.
\end{align}
And $\beta=1/k_{B}T$, $k_{B}$ is the Boltzmann constant. There are generally
three branches of valid solutions: (i) non-zero solution, $\Delta\neq0$ and
$\Xi\neq0$; (ii) zero solution, $\Delta=0$ and $\Xi\neq0$; (iii) zero
solution, $\Delta\neq0$ and $\Xi=0$. The one with the lowest energy is picked
out as the physically realized state. At zero temperature, the per site ground
energy has a simple form,
\end{subequations}
\begin{equation}
\frac{E_{0}}{N_{\Lambda}}=-\frac{3}{2}z\left(  J_{1}\Delta^{2}+\left\vert
J_{2}\right\vert \Xi^{2}\right)  .
\end{equation}

\subsection{Green's Function and Susceptibility}

In order to calculate the susceptibility we introduce the Matsubara Green's
function in the form of a $6\times6$ matrix,
\begin{equation}
G(k,\tau)=-\left\langle T_{\tau}\Phi_{k}\left(  \tau\right)  \Phi_{k}^{\dag
}\left(  0\right)  \right\rangle =\frac{1}{\beta}\sum_{n}G\left(
k,i\omega_{n}\right)  e^{-i\omega_{n}\tau}.
\end{equation}
The bosonic Matsubara Green's function $G\left(  k,i\omega_{n}\right)  $ is
generally worked out as,
\begin{equation}
G\left(  k,i\omega_{n}\right)  =\left(  i\omega_{n}\sigma_{z}\otimes
A_{0}-M_{k}\right)  ^{-1},
\end{equation}
where $\omega_{n}=2n\pi/\beta$. The three spectra, Eq.
(\ref{spectra BBM boson}), can also be read out from the poles of the Green's function.

As we shall study spin order as well as the nematic order in the system, we
define two types of correlation functions in Matsubara formalism. The first
type is the spin-spin correlation. Due to rotational invariance, we need only
to consider the imaginary-time spin-spin correlation for $S^{z}$,
\begin{equation}
\chi_{S^{z}}(q,\tau)=\left\langle T_{\tau}S^{z}(q,\tau)S^{z}%
(-q,0)\right\rangle .
\end{equation}
Its Fourier transform is given by%
\begin{equation}
\chi_{S^{z}}(q,i\omega_{n})=\int_{0}^{\beta}d\tau e^{i\omega_{n}\tau}%
\chi_{S^{z}}(q,\tau).
\end{equation}
The second type is the imaginary-time quadrupole-quadrupole correlation and
its Fourier transform defined for the quadrupole operators $Q$'s in Eq.
(\ref{quadrupole operators}) is given by,%
\begin{align}
\chi_{Q}(q,\tau) &  =\left\langle T_{\tau}Q(q,\tau)Q(-q,0)\right\rangle ,\\
\chi_{Q}(q,i\omega_{n}) &  =\int_{0}^{\beta}d\tau e^{i\omega_{n}\tau}\chi
_{Q}(q,\tau),\\
Q &  \in\{Q^{\left(  0\right)  },Q^{\left(  2\right)  },Q^{xy},Q^{yz}%
,Q^{zx}\}.
\end{align}
Due to the rotational invariance, here we only present two of them,
$\chi_{Q^{\left(  2\right)  }}$ and $\chi_{Q^{xy}}$ ($\chi_{Q^{\left(
0\right)  }}$ is equivalent to $\chi_{Q^{\left(  2\right)  }}$, $\chi_{Q^{yz}%
}$ and $\chi_{Q^{zx}}$ are equivalent to $\chi_{Q^{xy}}$). According to the
single-mode approximation theory \cite{Auerbach1994}, $\chi_{S^{z}}$ is
related to the spin order by the single mode $S^{z}(q)\left\vert
0\right\rangle \ $with spin density wave
\begin{equation}
S^{z}(q)=\sum_{i}e^{iq\cdot R_{i}}S_{i}^{\alpha},
\end{equation}
while $\chi_{Q}$ is related to the nematic order by the single mode
$Q(q)\left\vert 0\right\rangle \ $with quadrupole density wave
\begin{equation}
Q(q)=\sum_{i}e^{iq\cdot R_{i}}Q_{i}.
\end{equation}
The expressions of the susceptibilities at zero temperature can be found in
Appendix. A.

\section{Long-Range Spin Nematic Order on Triangular Lattice}

\begin{figure}
[ptb]
\begin{center}
\includegraphics[
height=1.9626in,
width=3.1631in
]%
{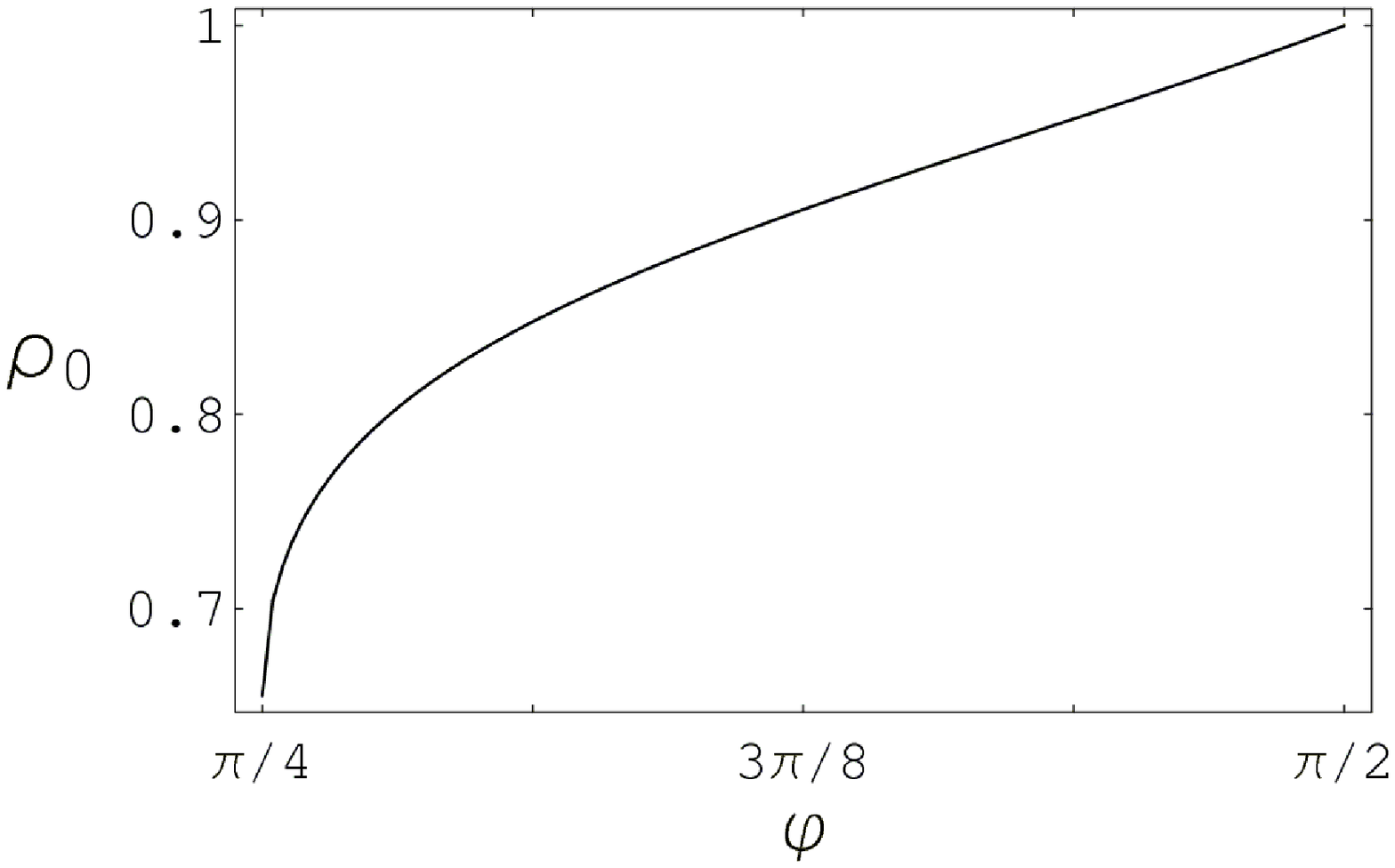}%
\caption{The condensation density $\rho_{0}$ at zero temperature.}%
\label{ro0vsphi}%
\end{center}
\end{figure}

The non-zero solution of the mean field parameters satisfies%
\begin{equation}
\widetilde{\Delta}+\widetilde{\Xi}=\frac{2}{3\sqrt{3}},
\end{equation}
at zero temperature and on the triangular lattice. With this relation, the
spectrum $\omega_{2}\left(  k\right)  $ becomes gapless at the point, $\ $%
\begin{equation}
k^{\ast}=\left(  k_{x}^{\ast},k_{y}^{\ast}\right)  =\left(  \frac{\pi}%
{3},\frac{\pi}{3}\right)  ,
\end{equation}
where the boson condensation occurs. As temperature becomes nonzero, the
spectrum $\omega_{2}\left(  k\right)  $ will open a gap and thus no
condensation occurs. When the condensation occurs, we should parse the
condensation terms and rewrite the equations as
\begin{subequations}
\begin{align}
\rho_{0} &  =2-\int\frac{d^{2}k}{\left(  2\pi\right)  ^{2}}\frac
{1-\widetilde{\Xi}\ \eta_{k}}{\widetilde{\omega}_{2}\left(  k\right)  },\\
\Delta &  =\frac{1}{2}\rho_{0}+\frac{1}{3\sqrt{3}}\int\frac{d^{2}k}{\left(
2\pi\right)  ^{2}}\frac{\widetilde{\Delta}\ \eta_{k}^{2}}{\widetilde{\omega
}_{2}\left(  k\right)  },\\
\Xi &  =\frac{1}{2}\rho_{0}+\frac{1}{3\sqrt{3}}\int\frac{d^{2}k}{\left(
2\pi\right)  ^{2}}\frac{\left(  1-\widetilde{\Xi}\ \eta_{k}\right)  \ \eta
_{k}}{\widetilde{\omega}_{2}\left(  k\right)  },
\end{align}
where the condensation density is (see numerical result in Fig. \ref{ro0vsphi}%
)
\end{subequations}
\begin{equation}
\rho_{0}=\left[  \frac{2n_{B}\left(  \widetilde{\omega}_{2}\left(  k^{\ast
}\right)  \right)  +1}{N_{\Lambda}\widetilde{\omega}_{2}\left(  k^{\ast
}\right)  }\right]  \frac{3\sqrt{3}\widetilde{\Delta}}{2}.\label{ro0}%
\end{equation}%

\begin{figure}
[ptbptb]
\begin{center}
\includegraphics[
height=1.7864in,
width=3.1718in
]%
{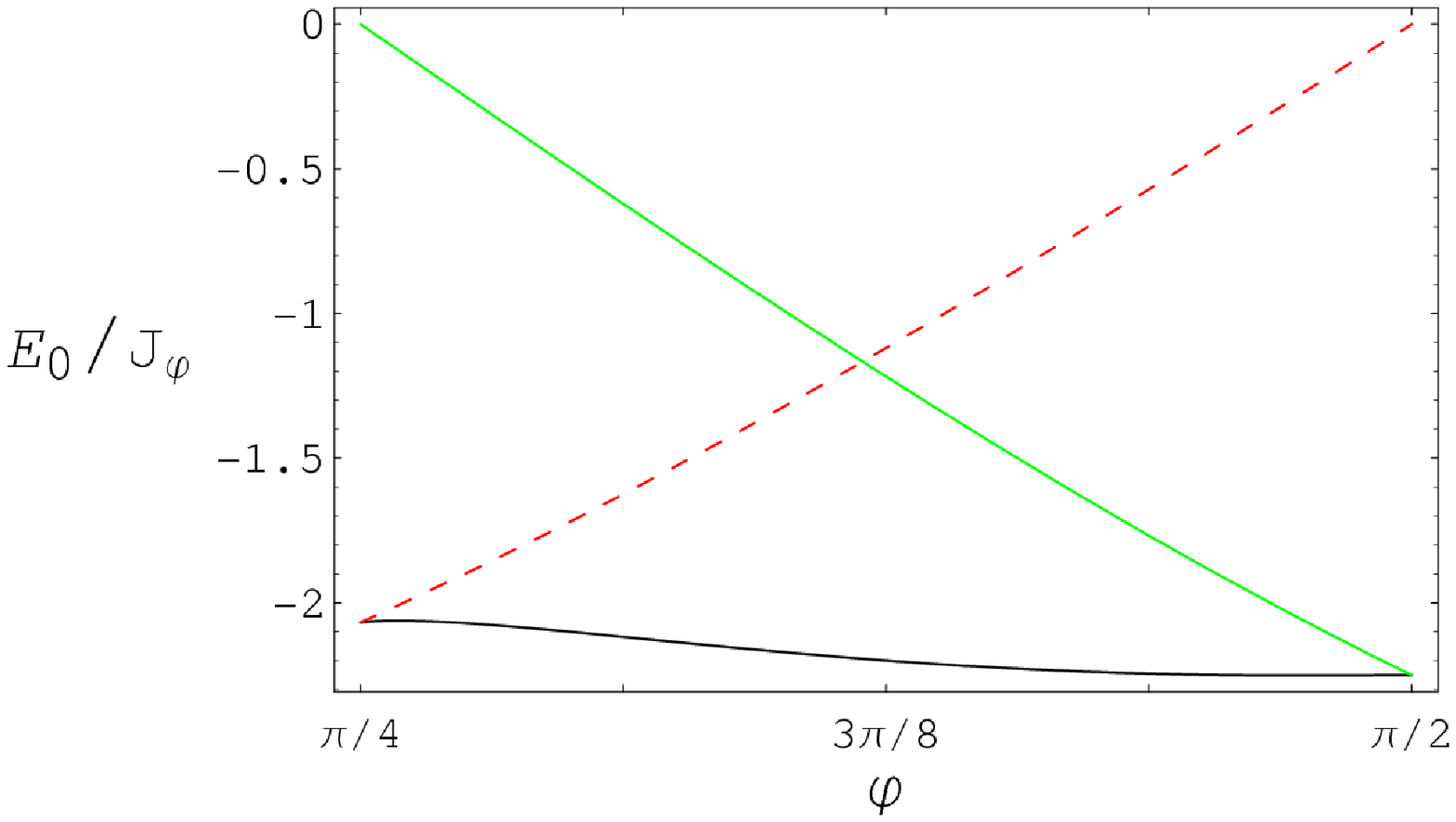}%
\caption{(Color online) The ground energies. The solid line is the non-zero
solution: $\Delta\neq0,\Xi\neq0$. The dashed line is the zero solution:
$\Delta\neq0,\Xi=0$. The dotted line is the zero solution: $\Delta=0,\Xi\neq
0$. This figure shows the non-zero solution is the optimized one.}%
\label{E0vsphi}%
\end{center}
\end{figure}

From the ground energies shown in Fig. \ref{E0vsphi},%

we see the non-zero solution is the optimized one in the range $\pi
/4<\varphi<\pi/2$. At the SU($3$) point of $\varphi=\pi/4$, the zero solution
with $\Delta\neq0$ and $\Xi=0$ is degenerate with the non-zero solution. At
the point of $\varphi=\pi/2$, the zero solution with $\Delta=0$ and $\Xi\neq0$
is degenerate with the non-zero solution. This reflects the fact that the two
points are highly symmetric points. In the range $\pi/4<\varphi<\pi/2$, the
condensate is non-zero. So what is the physical effect of the quasiparticle
condensation? By probing the possible orders in the system, we find the
condensation leads to the nematic LRO while spin moments vanish, i.e. the
nematic state is non-magnetic. This conclusion is drawn from the static spin
and quadrupole structure factors shown in Fig. \ref{Xall}. (Please refer to
the expressions listed in Appendix. A.)%

\begin{figure}
[ptbptbptb]
\begin{center}
\includegraphics[
trim=0.015932in -0.031719in -0.006437in 0.040003in,
height=10.6812cm,
width=8.1781cm
]%
{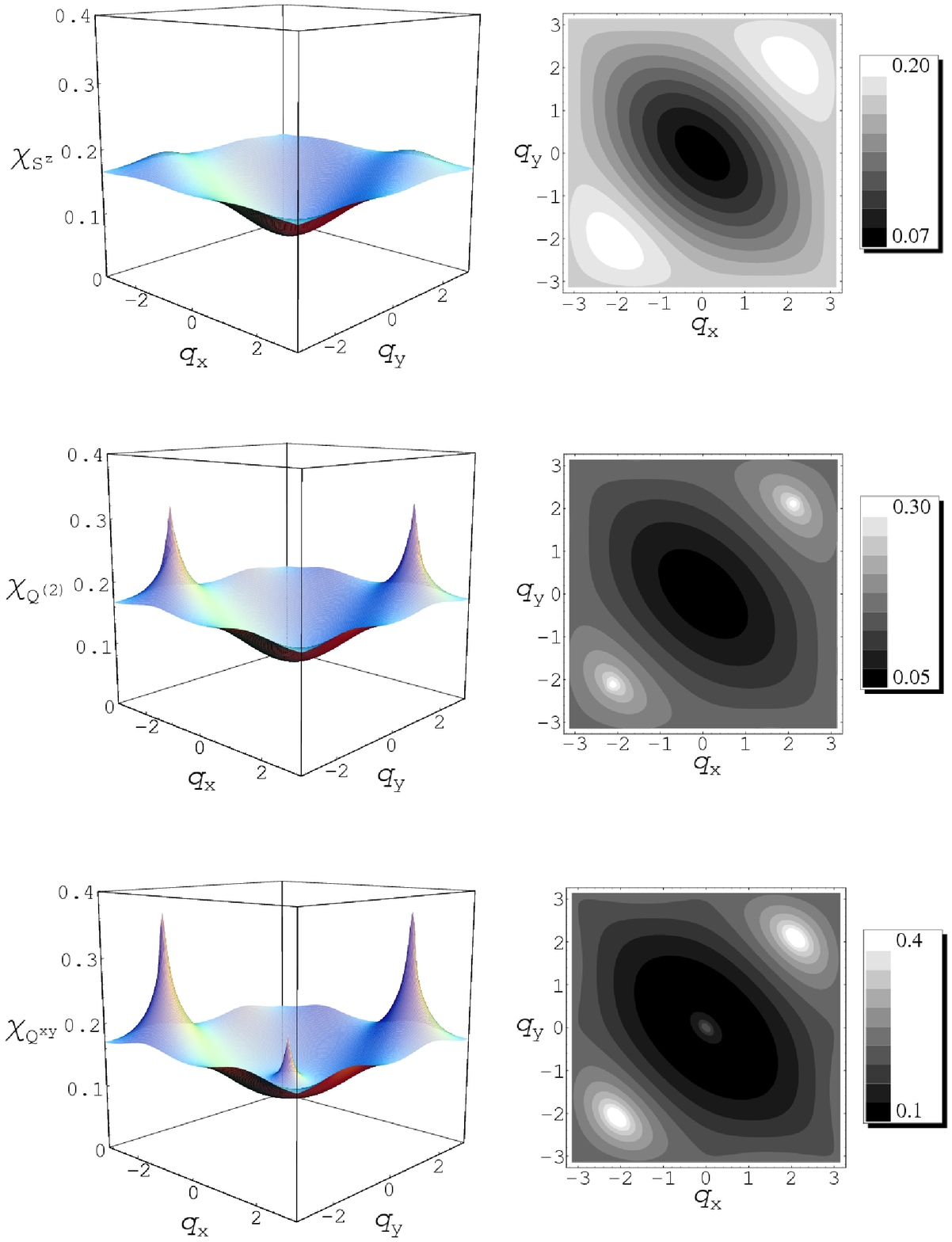}%
\caption{(Color online) Static structure factors $\chi_{S^{z}}$,
$\chi_{Q^{\left(  2\right)  }}$, and $\chi_{Q^{xy}}$ at a sampled point
$\varphi=0.863938$. The images at other point in the regime $\frac{\pi}%
{4}<\varphi<\frac{\pi}{2}$ are qualitatively the same.}%
\label{Xall}%
\end{center}
\end{figure}

\begin{figure}
[ptb]
\begin{center}
\includegraphics[
height=2.0838in,
width=3.1266in
]%
{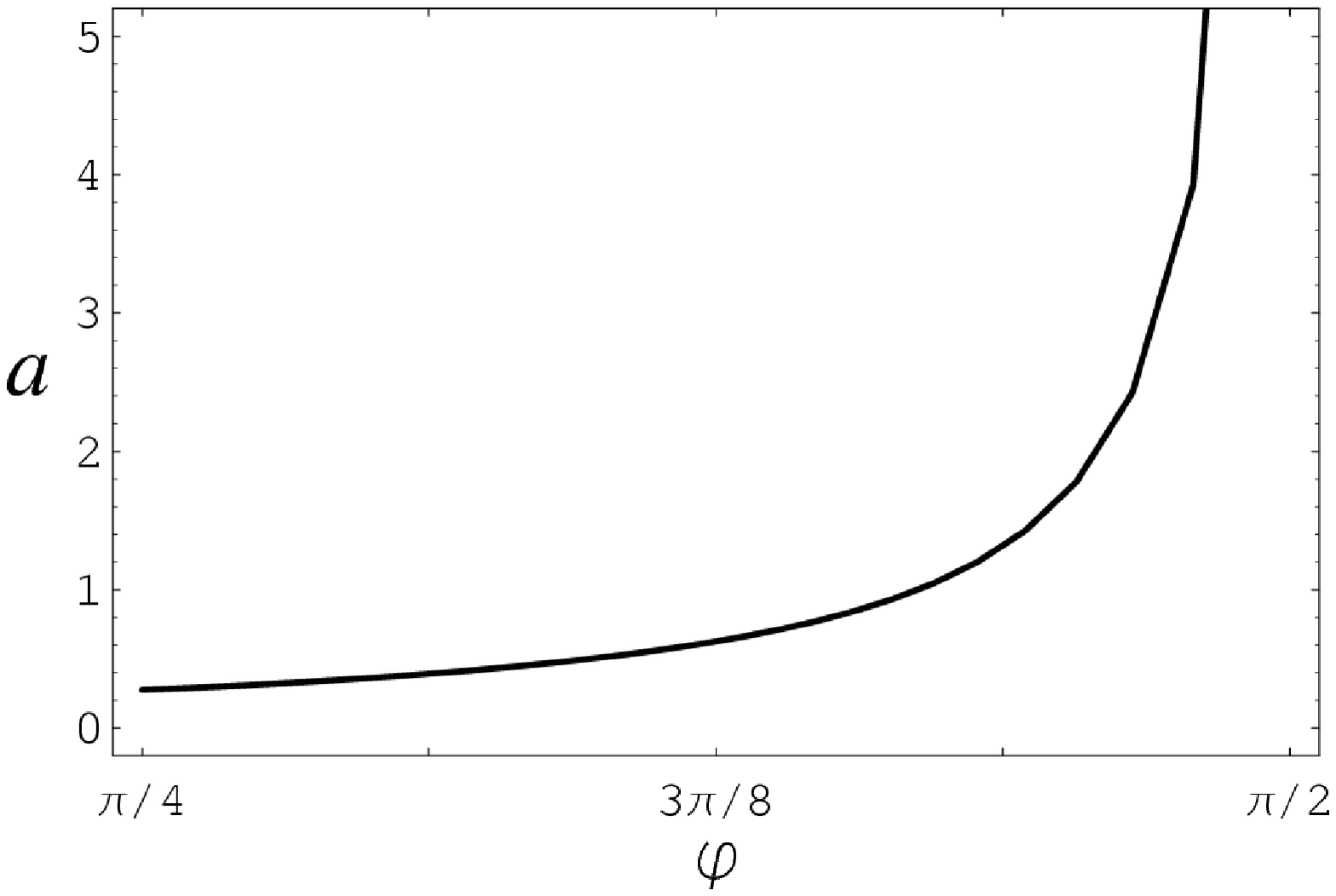}%
\caption{The coefficient $a$ in Eq. (\ref{linearDOS}).}%
\label{avsphi}%
\end{center}
\end{figure}

The static quadrupole structure factor $\chi_{Q^{\left(  2\right)  }}%
(q,\tau=0^{+})$ and $\chi_{Q^{xy}}(q,\tau=0^{+})$ show sharp divergent peaks
at the two points $q^{\ast}=\pm2k^{\ast}=\pm\left(  \frac{2\pi}{3},\frac{2\pi
}{3}\right)  $ indicating the existence of antiferro-quadrupole LRO. While the
static spin structure factor $\chi_{S^{z}}(q,\tau=0^{+})$ shows two small
humps at $q^{\ast}=\pm\left(  \frac{2\pi}{3},\frac{2\pi}{3}\right)  $.
Surprisingly $\chi_{Q^{xy}}(q,\tau=0^{+})$ also exhibits a divergent peak at
the point $q^{0}=\left(  0,0\right)  $. Analytically, at the condensate
points, the divergent terms of $\chi_{Q^{xy}}$ are parsed out as (obtained by
taking the limit $T\rightarrow0$ in the finite temperature results)
\begin{subequations}
\label{N0XQxy}%
\begin{align}
\chi_{Q^{xy}}^{\rho_{0}}(q^{0})  & =\frac{1}{9}\rho_{0}^{2}N_{\Lambda},\\
\chi_{Q^{xy}}^{\rho_{0}}(q^{\ast})  & =\frac{2}{9}\rho_{0}^{2}N_{\Lambda}.
\end{align}
The ratio of the weights of the ferro- and antiferro-quadrupole divergent
peaks is
\end{subequations}
\begin{equation}
r=\frac{\chi_{Q^{xy}}^{\rho_{0}}(q^{0})}{\chi_{Q^{xy}}^{\rho_{0}}(q^{\ast}%
)}=\frac{1}{2}.
\end{equation}
$\chi_{Q^{xy}}^{\rho_{0}}(q^{0})$ and $\chi_{Q^{xy}}^{\rho_{0}}(q^{\ast})$ are
proportional to the number of the lattice site $N_{\Lambda}$, which indicates
that the ferro- and antiferro-quadrupole LRO coexist for $Q_{i}^{xy}$. The
result can be understood well if one notices that the states defined in Eq.
(\ref{Qeigenstates}) are eigenstates of $Q_{i}^{\left(  2\right)  }$,
\begin{subequations}
\label{Q2eigenstates}%
\begin{align}
\left\vert Q_{i}^{\left(  2\right)  }=-1\right\rangle  &  =b_{i1}^{\dag
}\left\vert 0\right\rangle ,\\
\left\vert Q_{i}^{\left(  2\right)  }=1\right\rangle  &  =b_{i2}^{\dag
}\left\vert 0\right\rangle ,\\
\left\vert Q_{i}^{\left(  2\right)  }=0\right\rangle  &  =b_{i3}^{\dag
}\left\vert 0\right\rangle .
\end{align}
If on a bipartite lattice, $\left\vert Q_{i}^{\left(  2\right)  }%
=-1\right\rangle $ and $\left\vert Q_{i}^{\left(  2\right)  }=1\right\rangle $
would align in a staggered pattern with antiferro-quadrupole LRO in the
classical point of view. While on the triangular lattice, the LRO arrangement
of quadrupole moments, $\left\langle Q_{i}^{\left(  2\right)  }\right\rangle
$, is a 2$\pi/3$ structure as we revealed above. Nevertheless they are not the
eigenstates of $Q_{i}^{xy}$, instead
\end{subequations}
\begin{subequations}
\begin{align}
\left\vert Q_{i}^{\left(  2\right)  }=-1\right\rangle  & =\frac{1}{\sqrt{2}%
}\left(  \left\vert Q_{i}^{xy}=1\right\rangle -\left\vert Q_{i}^{xy}%
=-1\right\rangle \right)  ,\\
\left\vert Q_{i}^{\left(  2\right)  }=1\right\rangle  & =\frac{-1}{\sqrt{2}%
}\left(  \left\vert Q_{i}^{xy}=1\right\rangle +\left\vert Q_{i}^{xy}%
=-1\right\rangle \right)  .
\end{align}
So the antiferro-quadrupole moment of $Q_{i}^{\left(  2\right)  }$ means the
existence of both the ferro- and antiferro-quadrupole moments of $Q_{i}^{xy}$.
This phenomenon is reminiscent of the ferrimagnets or the canted
anteferromagnets, which exhibit both ferromagnetic and antiferromagnetic orders.

Now we address to the two terminals of the range $\pi/4\leqslant
\varphi\leqslant\pi/2$. At the SU($3$) point $\varphi=\pi/4$, the static spin
structure factor also becomes divergent at $q^{\ast}=\pm\left(  \frac{2\pi}%
{3},\frac{2\pi}{3}\right)  $. The state is degenerated with the nematic phase,
which reflects the higher symmetry of the system \cite{Chen1971,Chen1973}. At
$\varphi=\pi/2$ where the Hamiltonian becomes $H=\sum_{ij}(\mathbf{S}_{i}%
\cdot\mathbf{S}_{j})^{2}$, all of the static structure factors vanish and the
system is free of both spin and quadrupole moments, which means the system is
totally disordered.

The Matsubara formalism facilitates the evaluation of the expectation values
at finite temperatures ($\pi/4<\varphi<\pi/2$),
\end{subequations}
\begin{subequations}
\begin{align}
\left\langle b_{i1}^{\dag}b_{i1}\right\rangle  & =\left\langle b_{i2}^{\dag
}b_{i2}\right\rangle =\frac{1}{3},\\
\left\langle b_{i1}^{\dag}b_{i2}\right\rangle  & =\left\langle b_{i2}^{\dag
}b_{i1}\right\rangle =\frac{1}{6}-\frac{1}{2}n_{B}\left(  \lambda\right)  .
\end{align}
Thus we obtain
\end{subequations}
\begin{subequations}
\label{moments}%
\begin{align}
\left\langle S_{i}^{z}\right\rangle  & =0,\\
\left\langle Q_{i}^{\left(  2\right)  }\right\rangle  & =0,\\
\left\langle Q_{i}^{xy}\right\rangle  & =-\left[  \frac{1}{3}-n_{B}\left(
\lambda\right)  \right]  .\label{moments3}%
\end{align}
The uniform quadrupole moment $\left\langle Q_{i}^{xy}\right\rangle $ keeps
nonzero at finite temperatures. This result does not mean the
finite-temperature phase transition, just shows a robust uniform quadrupole moment.

It is well known, for the noninteracting Bose gas in three dimensions, the
density of states (DOS) will vanish with the energy decreasing down to zero,
i.e. $D\left(  E\right)  \sim\sqrt{E}\rightarrow0$ when $E\rightarrow0$, which
leads to the Bose condensation at the gapless point $E=0$ at low temperatures.
The situation is similar here. The DOS is defined by
\end{subequations}
\begin{equation}
D\left(  E\right)  =\sum_{k,\mu}\delta\left(  E-\omega_{\mu}\left(  k\right)
\right)  .
\end{equation}
Since there exist a flat band $\omega_{1}=\lambda$ for quasi-particles,
$D(E)$\ always has a divergent peak at $E=\lambda$. We find that DOS rises
linearly in $E$ from zero in the range of $\pi/4\leqslant\varphi<\pi/2$,
\begin{equation}
D\left(  E\right)  \sim a\left(  \varphi\right)  E+O(E^{2}),\label{linearDOS}%
\end{equation}
because the gapless spectrum $\omega_{2}\left(  k\right)  $ exhibits a node at
$k^{\ast}=\left(  \frac{\pi}{3},\frac{\pi}{3}\right)  $. By the DOS in Eq.
(\ref{linearDOS}), the low temperature specific heat is shown to exhibit the
law of $T^{2}$,%
\begin{align}
C_{V}  & =\int dE\ D\left(  E\right)  \frac{E}{2k_{B}T}\left(  \sinh\frac
{E}{2k_{B}T}\right)  ^{-2}\nonumber\\
& \sim\frac{2\pi^{2}}{3}a\left(  \varphi\right)  k_{B}^{2}T^{2}.\label{Cv}%
\end{align}
The coefficient $a\left(  \varphi\right)  $ is plotted in Fig. \ref{avsphi}.%

While at the terminal $\varphi=\pi/2$, the node of the spectra disappears and
the DOS has the form $D\left(  E\right)  \sim b+cE$ with $b\neq0$, then one
would get $C_{V}\sim\frac{\pi^{2}}{3}bk_{B}T$.

It is noteworthy that, at nonzero temperatures, the spectrum $\omega
_{2}\left(  k\right)  $ is gapful and the DOS always has $D\left(  E\right)
\sim b+cE$. But at very low temperatures, $b$ is quite small and the power law
in Eq. (\ref{Cv}) can be satisfied asymptotically.

\begin{figure}
[ptb]
\begin{center}
\includegraphics[
height=1.8796in,
width=3.1158in
]%
{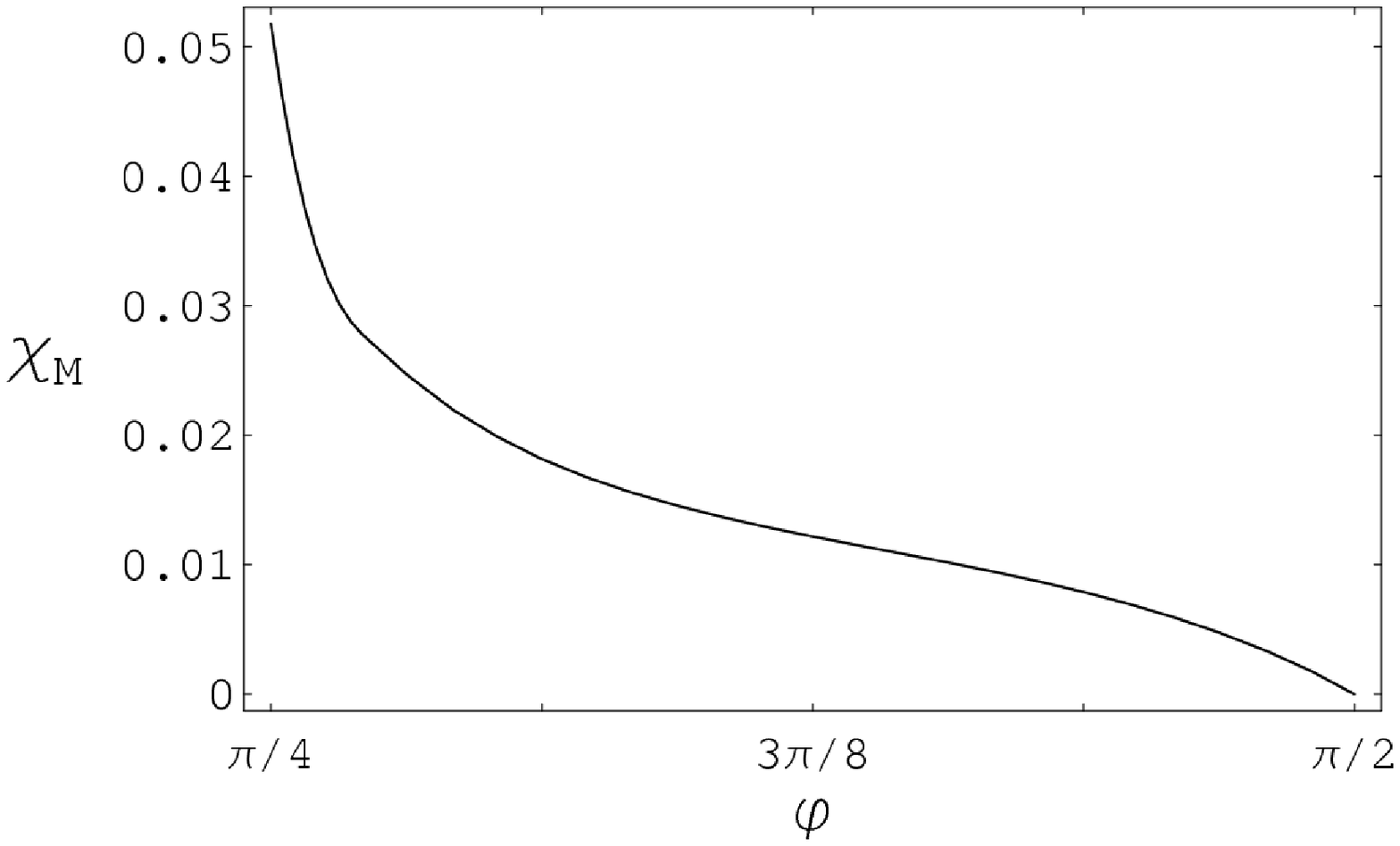}%
\caption{The magnetic susceptibility $\chi_{M}$ at zero temperature.}%
\label{XMvsphi}%
\end{center}
\end{figure}

The uniform magnetic susceptibility at zero temperature is obtained by
Kramers-Kronig relation \cite{Auerbach1994}%
\begin{equation}
\chi_{M}=\lim_{q\rightarrow0}\frac{1}{\pi}\int_{0}^{\infty}d\omega
\frac{\operatorname{Im}\chi_{S^{z}}(q,\omega)}{\omega}.
\end{equation}
or by analytic continuation \cite{Mahan}
\begin{equation}
\chi_{M}=\lim_{q\rightarrow0}\lim_{i\omega_{n}\rightarrow0}\chi_{S^{z}%
}(q,i\omega_{n}),
\end{equation}
They give the same result. At zero temperature, $\chi_{M}$ versus $\varphi$ is
illustrated in Fig. \ref{XMvsphi}.%

$\chi_{M}$ reaches the maximal value at the SU($3$) point $\varphi=\pi/4$,
while it approaches zero at the end point $\varphi=\pi/2$ where the system is
inert to the external inspiration.

\section{Discussion}

Recently the insulating antiferromagnet NiGa$_{2}$S$_{4}$ arouses much
attention \cite{Nakatsuji2005}. The spin disorder observed in the experiment
suggest that it may be a realization of the conceptual spin liquid that has
long been explored over the past decades. This chalcogenide has a stacked
triangular lattice with weak interactions between layers. Strong Hund's
coupling in Ni$^{2+}$($t_{2g}^{6}e_{g}^{2}$) leads to the magnetism with spin
$S=1$. Magnetic neutron scattering shows absence of conventional magnetic
order and excludes the possibility of bulk spin glass freezing at low
temperatures. Its specific heat shows low temperature power law, $C_{V}\sim
T^{2}$, indicating gapless excitations and linearly dispersive modes in two
dimensions. No divergence was observed for the magnetic susceptibility with
the temperature decreasing down to $0.35K$. These features can be produced by
the nematic state as we studied above. However the incommensurate short-range
order observed in the experiment still remains untouched.

Our bond-operator mean-field theory has the same origin as the SBMFT and is
superior to the molecular-field approximation (MFA) because the MFA starts
from a prescribed ferro- or antiferro-order and produces the same result
regardless of the dimensionality \cite{Chen1973,Harada02}, while the our
theory has no bias on the order or disorder of the ground state in advance. In
two dimensions, we got a gapless nematic phase with quadruple LRO as we
illustrated above for the triangular lattice. In one dimension, we got a
gapped nematic phase \cite{LiP2006}. Nevertheless, any mean-field theory can
not be conclusive in its own right, thus other methods for the same problem
are demanded. Unlike the spin wave theory, the bond-operator mean-field theory
used in this paper does not prescribe an ordered state in advance. It has the
same origin as SBMFT, except the species of bosons is altered from two to
three. The antiferro-quadruple LRO emerges as a consequence of the
condensation of the SU($3$) bosons. Our results also show the quadruple
operators are divided into two types. They have different LRO patterns and
should be considered differently in a spin wave theory (i.e., for $Q^{(0)}$
and $Q^{(2)}$, one need only to consider their antiferro-orders; while for
$Q^{xy}$, $Q^{yz}$, and $Q^{zx}$, one should consider their ferro- and
antiferro-orders at the same time. The coexistence of ferro- and
antiferro-quadruple LRO's reveals that the quadruple operators can not be
considered as the analogues of spin operators in the magnetic LRO phenomena.
And as a merit of bosonic language, we expect that this theory gives good
estimation of ground energy values (see Fig. \ref{E0vsphi}), like the SBMFT
\cite{LiP2005,Ceccatto}. To see how this theory describes the
antiferromagnetic phase ($-\pi/4<\varphi<\pi/4$) of the SBBM is also
desirable, which will be considered in our future work.

In summary, a SU($3$) bosons representation is introduced and the associated
bond-operator mean-field theory is established to describe the
antiferro-nematic phase of SBBM. It is revealed delicately that this nematic
state may exhibit both the ferro- and antiferro-quadruple LRO's, which is
reminiscent of the ferrimagnets or the canted antiferromagnets. And the
quadruple LRO may survive to finite temperatures. The system may provide a
rare example exhibiting finite-temperature phase transition in two dimensions.
Possible relevance of this unconventional state to the quasi-two-dimensional
triangular material NiGa$_{2}$S$_{4}$ is addressed.

This work was supported by the Research Grant Council of Hong Kong under Grant
No.: HKU7038/04P.

\appendix{}

\section{Static spin and quadrupole structure factors}

The static spin and quadrupole structure factors at zero temperature,
$\chi(q,\tau=0^{+})$, are worked out as
\begin{subequations}
\begin{align}
\chi_{S^{z}}(q,0^{+})  & =\int\frac{d^{2}k}{\left(  2\pi\right)  ^{2}}%
[A_{2}\left(  k\right)  +A_{3}\left(  k\right) \nonumber\\
& +3A_{2}\left(  k\right)  B_{2}\left(  k+q\right)  +3A_{3}\left(  k\right)
B_{3}\left(  k+q\right) \nonumber\\
& +3C_{2}\left(  k\right)  C_{2}\left(  k+q\right)  +3C_{3}\left(  k\right)
C_{3}\left(  k+q\right)  ],\nonumber\\
& \\
\chi_{Q^{\left(  2\right)  }}(q,0^{+})  & =\int\frac{d^{2}k}{\left(
2\pi\right)  ^{2}}[A_{2}\left(  k\right)  +A_{3}\left(  k\right) \nonumber\\
& +3A_{2}\left(  k\right)  B_{3}\left(  k+q\right)  +3A_{3}\left(  k\right)
B_{2}\left(  k+q\right) \nonumber\\
& +3C_{2}\left(  k\right)  C_{3}\left(  k+q\right)  +3C_{3}\left(  k\right)
C_{2}\left(  k+q\right)  ],\nonumber\\
& \\
\chi_{Q^{xy}}(q,0^{+})  & =\int\frac{d^{2}k}{\left(  2\pi\right)  ^{2}}%
[\frac{1}{3}A_{2}\left(  k\right)  +\frac{1}{3}A_{3}\left(  k\right)
\nonumber\\
& +A_{2}\left(  k\right)  B_{2}\left(  k+q\right)  +A_{3}\left(  k\right)
B_{3}\left(  k+q\right) \nonumber\\
& +4A_{2}\left(  k\right)  B_{3}\left(  k+q\right)  +4A_{3}\left(  k\right)
B_{2}\left(  k+q\right) \nonumber\\
& +4C_{2}\left(  k\right)  C_{3}\left(  k+q\right)  +4C_{3}\left(  k\right)
C_{2}\left(  k+q\right) \nonumber\\
& -C_{2}\left(  k\right)  C_{2}\left(  k+q\right)  -C_{3}\left(  k\right)
C_{3}\left(  k+q\right)  ],\nonumber\\
&
\end{align}
where the abbreviated notations are
\end{subequations}
\begin{subequations}
\begin{align}
A_{2}\left(  k\right)   & =\frac{1}{6}\left[  1+\frac{\lambda-\sqrt{3}\Xi_{k}%
}{\omega_{2}\left(  k\right)  }\right]  ,\\
A_{3}\left(  k\right)   & =\frac{1}{6}\left[  1+\frac{\lambda+\sqrt{3}\Xi_{k}%
}{\omega_{3}\left(  k\right)  }\right]  =A_{2}\left(  -k\right)  ,\\
B_{2}\left(  k\right)   & =\frac{1}{6}\left[  1-\frac{\lambda-\sqrt{3}\Xi_{k}%
}{\omega_{2}\left(  k\right)  }\right]  ,\\
B_{3}\left(  k\right)   & =\frac{1}{6}\left[  1-\frac{\lambda+\sqrt{3}\Xi_{k}%
}{\omega_{3}\left(  k\right)  }\right]  =B_{2}\left(  -k\right)  ,\\
C_{2}\left(  k\right)   & =\frac{1}{6}\frac{\sqrt{3}\Delta_{k}}{\omega
_{2}\left(  k\right)  },\\
C_{3}\left(  k\right)   & =\frac{1}{6}\frac{\sqrt{3}\Delta_{k}}{\omega
_{3}\left(  k\right)  }=-C_{2}\left(  -k\right)  .
\end{align}
$\Xi_{k}$ and $\Delta_{k}$ can be found in Eq. (\ref{DeltaCheta}). Note that
for $\chi_{Q^{\left(  2\right)  }}(q,0^{+})$ and $\chi_{Q^{xy}}(q,0^{+}) $,
one should consider the divergent terms at the points $q^{\ast}=\pm\left(
\frac{2\pi}{3},\frac{2\pi}{3}\right)  $ and $q^{0}=\left(  0,0\right)  $, such
as Eq. (\ref{N0XQxy}).
\end{subequations}

\end{document}